\documentclass{PoS}

\title{Project RISARD}

\ShortTitle{Project RISARD}

\author{\speaker{Marcin P. Gawro\'nski}, Krzysztof Go\'zdziewski, Krzysztof Katarzy\'nski\\
        Centre for Astronomy, Faculty of Physics, Astronomy and Informatics, 
        Nicolaus Copernicus University, Grudziadzka 5, 87-100 Toru\'n, Poland\\
        E-mail: \email{motylek@astro.umk.pl}$^*$}

\abstract{Red Dwarfs (RDs) are the most common, low-mass stars ($\lesssim$\,0.5\,M$_{\odot}$) in the Solar neighbourhood,
and probably in the Universe as well. Most (likely all) young RDs are magnetically active, and
therefore it is impossible to measure their radial velocities (RVs) with the accuracy required by
contemporary planetary searches ( 5-50 m/s). Some of RDs are known as a source of variable
radio emission at centimetre wavelengths. This radiation is relatively weak ($\sim$\,0.2 to 1 mJy).
However, recent development of the EVN and e-VLBI systems in terms of sensitivity gives a new
possibility to investigate such stars. Here, we present first results from our ongoing RISARD
project (Radio Interferometric Survey of Active Red Dwarfs). The main goal of this project is to
detect an exoplanet by direct, precise measurements of a RD position and possible changes to this
position caused by the planet.}

\FullConference{11th European VLBI Network Symposium \& Users Meeting,\\
		October 9-12, 2012\\
		Bordeaux, France}

\begin{document}

\section{Introduction}
It is believed that Red Dwarfs (RDs) are the most numerous stars in the Universe (>70\%).
These low-mass ($\lesssim$\,0.5M$_{\odot}$) objects constitute also more than 40\% of total stellar 
mas in our Galaxy \cite{hen98}. However, it is still unclear what is the formation rate of planets (or low-mass companions)
around these very common object. Is this rate similar, less or higher to the rate that we observe for
solar-type FGK stars? RDs with their low-masses are very attractive targets for planetary search
because in principle it should be possible to detect their companions down to the super-Earth masses.

The recent and ongoing RV surveys are focused on RDs samples that are necessarily biased
towards chromospherically quiet and old objects. For instance, the RV survey of M-dwarfs with the
VLT+UVES already includes a brown dwarf desert object and several low-mass companion candidates
around early-type M-dwarfs \cite{kur09}. A remarkable multi-planet resonant system was discovered
around the M-dwarf Gliese 876, hosting two Jupiters \cite{mar98}. The M-dwarfs are also intensively monitored
by the Geneva Planet Search Team (e.g. \cite{may09}) that reported a multiple system of six planets
around Gliese 581.

\begin{figure*}
\centering
\includegraphics[scale=0.97]{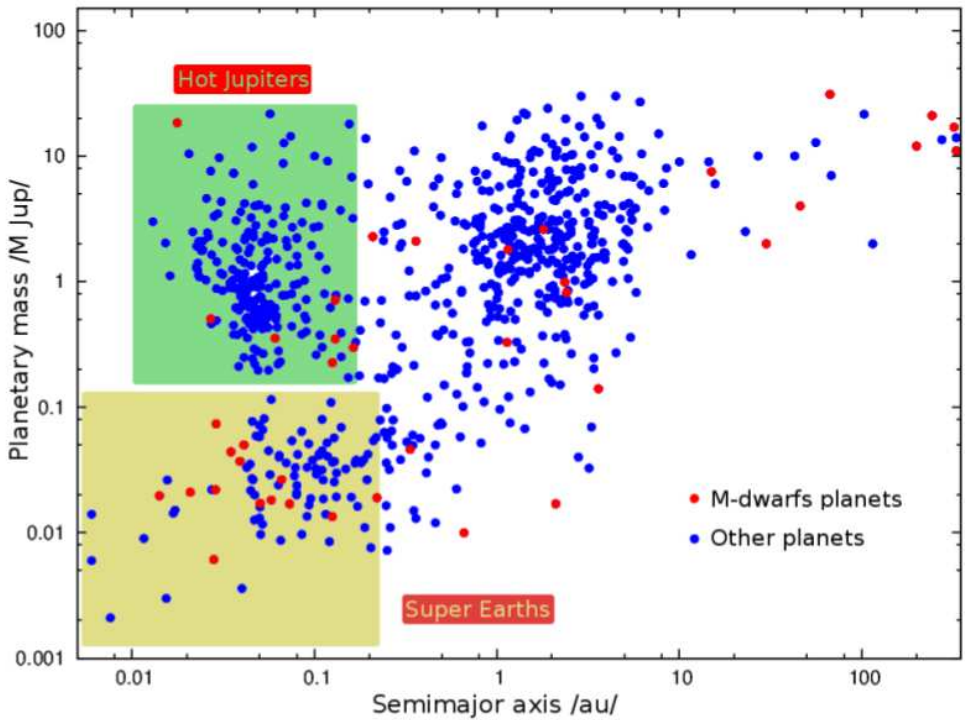}
\caption{Planetary systems statistics based on The Extrasolar Planets Encyclopaedia (http://exoplanet.eu/).
The M-dwarf planets are marked with red dots, the other planets with blue dots. An excess of Super-Earths
in the case of RDs in comparison to the other stars is clearly visible.}
\label{fig1}
\end{figure*}

The currently known mass function of planetary candidates around RDs suggests that their
jovian companions are less frequent than in the neighbourhood of the FGK-type stars. Especially the
lack of so-called hot Jupiters is very striking. However, planets in the Neptune-mass range may be
much more common (see Fig. \ref{fig1}). This is indicated by the recent discoveries from the RV surveys
(e.g. \cite{may09}) of many multiple super-Earth systems around quiet M-dwarfs. Also, regarding solar-like
dwarfs, the KEPLER transit detections indicate an exponential mass function of short-period
planets in that mass range. This statistic is biased towards an old stars and it has been suggested
\cite{joh09} that this could be affected by the expected correlation between the occurrence rate of planetary
systems and the metallicity of host stars \cite{san04}. It has been also pointed out that M-dwarfs hosting
planets appear systematically metal-rich \cite{joh09}.

Magnetically active young M-dwarfs are producing a weak, variable radio emission at radio
wavelengths \cite{gud93}. It is believed that the electron cyclotron maser instability \cite{hal08} 
and/or gyromagnetic emission (e.g. \cite{gud02}) are responsible for this radiation. The VLBI technique has been already
successfully used for the observations of low-mass active M-dwarfs \cite{pes00} or even brown dwarfs \cite{for09}.
The recent development of the EVN and e-VLBI systems improved significantly the sensitivity
and opened a new observational window for stellar astrophysics. Current capabilities of European
VLBI Network allow for observations of very weak ($\sim100\mu$Jy), compact radio sources with the
brightness temperature in the range of 10$^6$--10$^7$ K like active RDs. This gives a new opportunity
for planetary search in terms of an independent and very precise astrometric surveys.

\section{Project RISARD}
RISARD is a new astrometric project with EVN started in 2009. This project is dedicated for
observations of very young ($\lesssim$\,1 Gyr) low-mass stars. The selection of young targets increases the
probability of planetary system occurrence because of increased metallicity \cite{san04}. We attempt to
reach a region in the physical and parameter space that cannot be explored by the RVs or transit
photometry (due to orbital orientation bias and/or chromospheric activity). Moreover, the astrometric
method may detect the stellar wobble around the barycentre independently of the orbital
inclination. As the final outcome, we expect to obtain non-biased statistics of low-mass companions
around young M-dwarfs, and to estimate their occurrence rate. The mass function in the Solar
neighbourhood exhibits a curious brown dwarf (BD) desert. Recent discoveries suggest that BDs
may tend to form close to the low-mass stars. The astrometric signal of BDs should be easily detectable.
Hence, even a lack of detections may contribute to the statistics of the sample of active,
young RDs by setting upper limits to the masses of their putative companions. We stress that our
survey complements ongoing RIPL survey with the help of the VLBA \cite{bow09}. RIPL is a program
conducted at 4 cm wavelength, initiated in 2007 that is focused on 29 active RDs within distance
$\sim$\,10 pc from Sun.

\begin{figure*}
\centering
\includegraphics[scale=1.0]{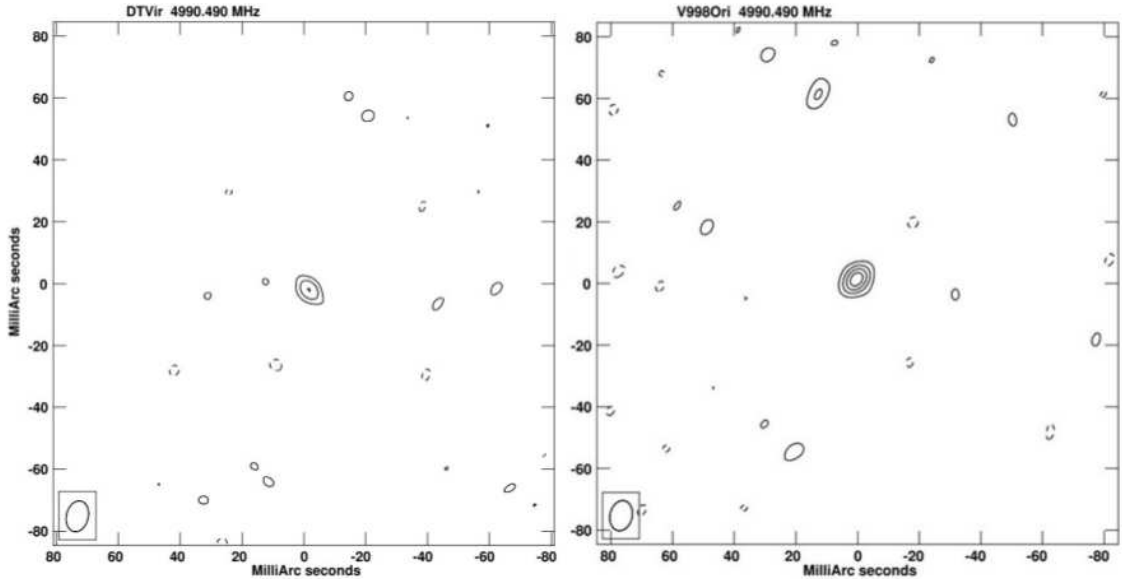}
\caption{Radio maps of two selected active RDs. \emph{Left:} DT\,Vir, \emph{right:} V998\,Ori.}
\label{fig2}
\end{figure*}

\section{Observations}
In the past three years, we have carefully prepared two test experiments to check the astrometric
capabilities of the western EVN. First, we tested the sensitivity level and the astrometric
accuracy (EG046, 7 hrs) and after that we conducted 48 hrs of e-VLBI observations (EG053) to
select a new sample of radio active RDs within the distance 10 $\lesssim d \lesssim$ 15 pc from Sun\footnote{The distance 
range has been chosen in order to not overlap with the RIPL sample.}. The
test observations EG046 has fully confirmed our expectation of the relative astrometric accuracy
$\sim$0.1--\,0.3 mas. In the experiment EG053 we observed 17 single, wide binary and multiple active
red and brown dwarfs. The expected level of the radio emission of these stars was $\gtrsim\,50\,\mu$Jy. We
estimated this level from the X-ray--radio luminosity correlation (e.g. \cite{ber10}). In addition we focused
on the objects with declination $\gtrsim$\,0. Preliminary results show the we have detected 12 RDs from
our initial EG053 sample. In the Fig. \ref{fig2} we show radio maps of two detected RDs.

Due to a proximity of selected stars to Sun, their high proper motions (0.1--0.4 mas/hr) and
physical sizes ($\lesssim$0.5\,mas) constitute a challenge to accurate astrometry. The observations of a
given target should not last longer than 1.5 hrs. Our experience gained thanks to the initial EG046
and EG053 phase shows that $\sim$2 hrs observation with phase-referencing (1.5 hrs on the target +
secondary calibrator) per target and 1 Gbps recording is sufficient to achieve an RMS at the level
of $\simeq$\,25\,mJy/beam and hence avoid any problems caused by proper motions of stars. During 2012
we have started regular astrometric observations of 12 RDs left in our sample. Second phase of
RISARD consisted of 3-4 epochs of observations for each target. Observations are finished and
data are currently reduced. After that phase we will re-evaluate again our sample and for the final
3-year part of RISARD only RDs with detection rate $\gtrsim$\,75\% will remain in the sample.

Based on the initial results we have simulated astrometric observations for the purpose of
RISARD project. Our calculations show that planets within the Jupiter-Saturn mass range will be
detectable, and a favourable sampling of the relative orbit would make it possible to determine the
Keplerian elements already from 10 observational epochs (note that in this example, we assume
the astrometric parameters pre-determined). An example simulation for one of stars from RISARD
sample is illustrated in Fig. 3. We present there also detection limits for putative low-mass components
for another our target. We plan to extend RISARD survey in the future and estimate the
dynamical masses of active RDs in binary systems placed within 20 pc from Sun.

\begin{figure*}
\centering
\includegraphics[scale=1.01]{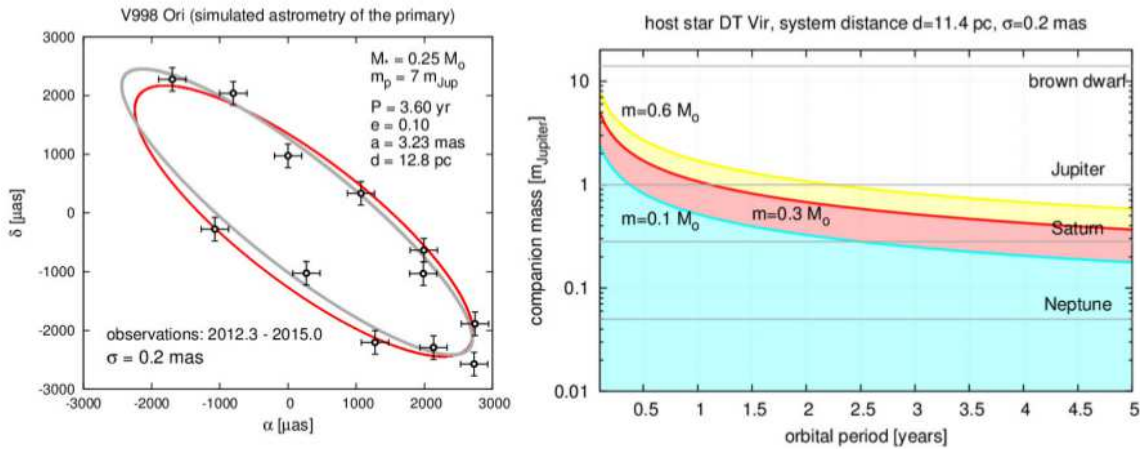}
\caption{Simulations of observations and the detection limits based on results from experiments EG046 and
EG053. \emph{Left:} A simulation of an astrometric signal from 7M$_{Jup}$ planet with the 3.6\,yr orbital period around
V998\,Ori. The mass of star has been assumed 0.25M$_{\odot}$, the distance 12.8 pc and the eccentricity of orbit
$e$ = 0.01. Red curve represents real orbit, grey -- orbit estimated with the use of the simulated observations.
\emph{Right:} The estimated detection limits for DT\,Vir. Calculations have been done for three different dynamical
masses of DT\,Vir. For the mass of host star in the range 0.1--0.3M$_{\odot}$ RISARD will be able to detect planets
in the Saturn-mass range on the > 4 yrs orbit. In both simulation the error of relative astrometry has been
assumed $\sigma$ = 0.2 mas.}
\label{fig3}
\end{figure*}

\newpage
\noindent
{\bf Acknowledgement}\\
\noindent

We are grateful to Polish National Science Centre for their support of project RISARD (grant
no. 2011/01/D/ST9/00735). The EVN is a joint facility of European, Chinese, South African, and
other radio astronomy institutes funded by their national research councils.

\end{document}